\documentclass[11pt]{article}
\usepackage{authblk}
\usepackage{amsmath, amssymb, graphics, setspace}
\usepackage[pdftex]{color,graphicx}
\usepackage[superscript]{cite}
\usepackage{verbatim}
\usepackage{multicol}
\usepackage[margin=1.0in]{geometry}
\usepackage{setspace}
\newcommand{\mathsym}[1]{{}}
\newcommand{\unicode}[1]{{}}

\begin{document}
\title{Quantum-enhanced sensing of a mechanical oscillator}
\author[1,2]{Katherine C. McCormick}
\author[1]{Jonas Keller}
\author[1,2]{Shaun C. Burd}
\author[1,2,3]{David J. Wineland}
\author[1]{Andrew C. Wilson} 
\author[1]{Dietrich Leibfried}
\affil[1]{National Institute of Standards and Technology, Boulder, CO 80305, USA}
\affil[2]{University of Colorado, Department of Physics, Boulder, CO 80305, USA}
\affil[3]{University of Oregon, Department of Physics, Eugene, OR 97403, USA}
\date{}
\maketitle
\begin{figure}
\centering
\includegraphics[width = \textwidth]{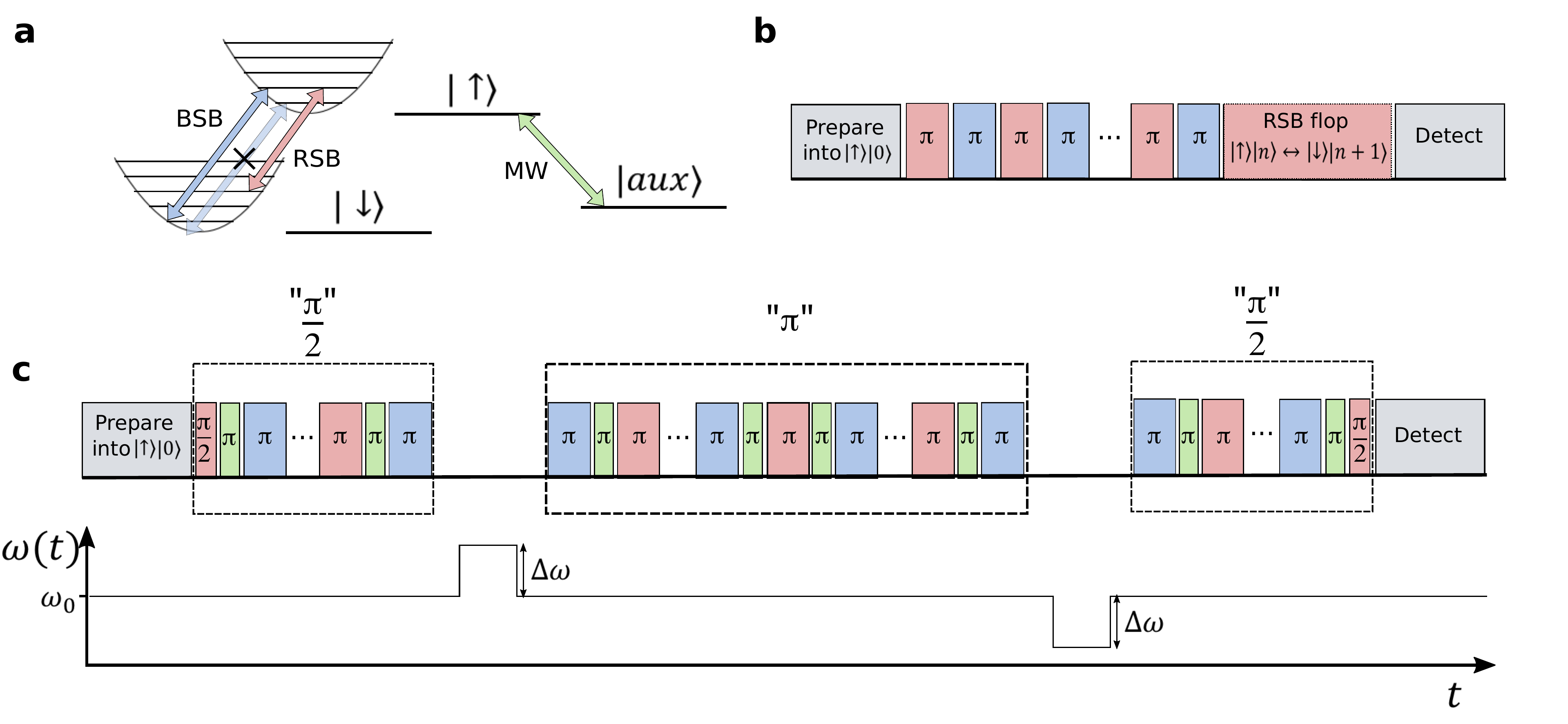}
\caption{\label{Fig:numberstate} {\bf Generating number states and number-state superpositions.} {\bf a,} Relevant energy levels and transitions for motional state creation. Blue sideband (BSB) pulses transfer population between $\lvert \downarrow\rangle\lvert k\rangle$ and $\lvert \uparrow\rangle\lvert k+1 \rangle$, while red sideband (RSB) pulses transfer population between $\lvert \uparrow\rangle\lvert k\rangle$ and $\lvert \downarrow\rangle\lvert k+1 \rangle$. The blue sideband does not couple to $\lvert \uparrow\rangle\lvert 0\rangle$ (shown as crossed out, faded blue arrow) because there is no resonant energy level below the ground state in the $\lvert \downarrow\rangle$ manifold. Transitions between the states $\lvert \uparrow \rangle$ and $\lvert aux\rangle$ are driven by a microwave transition (MW, indicated in green, which does not change $k$). {\bf b,} Pulse sequence for generating a pure harmonic oscillator number state. A series of alternating RSB and BSB $\pi$-pulses are applied, with each pulse adding one quantum of motion (or more quanta on higher order sidebands, see text) and flipping the spin of the state. To analyze the resulting state, a final RSB pulse is applied for a variable duration (labeled ``RSB flop") and the final spin state is detected via state-selective fluorescence. {\bf c,} Pulse sequences and trap frequencies for number-state interferometers. The first effective $\pi/2$-pulse (labeled ``$\pi/2$'') creates $\lvert 0\rangle + \lvert n\rangle$, followed by a free-precession period during which the mode frequency is increased by $\Delta \omega$. An effective $\pi$-pulse (``$\pi$'') swaps the phase of the superposition according to $\lvert 0\rangle+e^{i \phi}\lvert n\rangle \rightarrow e^{i \phi}\lvert 0\rangle+ \lvert n\rangle$. After another free-precession period with the mode frequency reduced by $\Delta \omega$, a final effective $\pi/2$-pulse closes the interferometer. For details on the composition of the effective pulses see text.}
\end{figure}
\begin{figure}
\centering
\includegraphics[width = \textwidth]{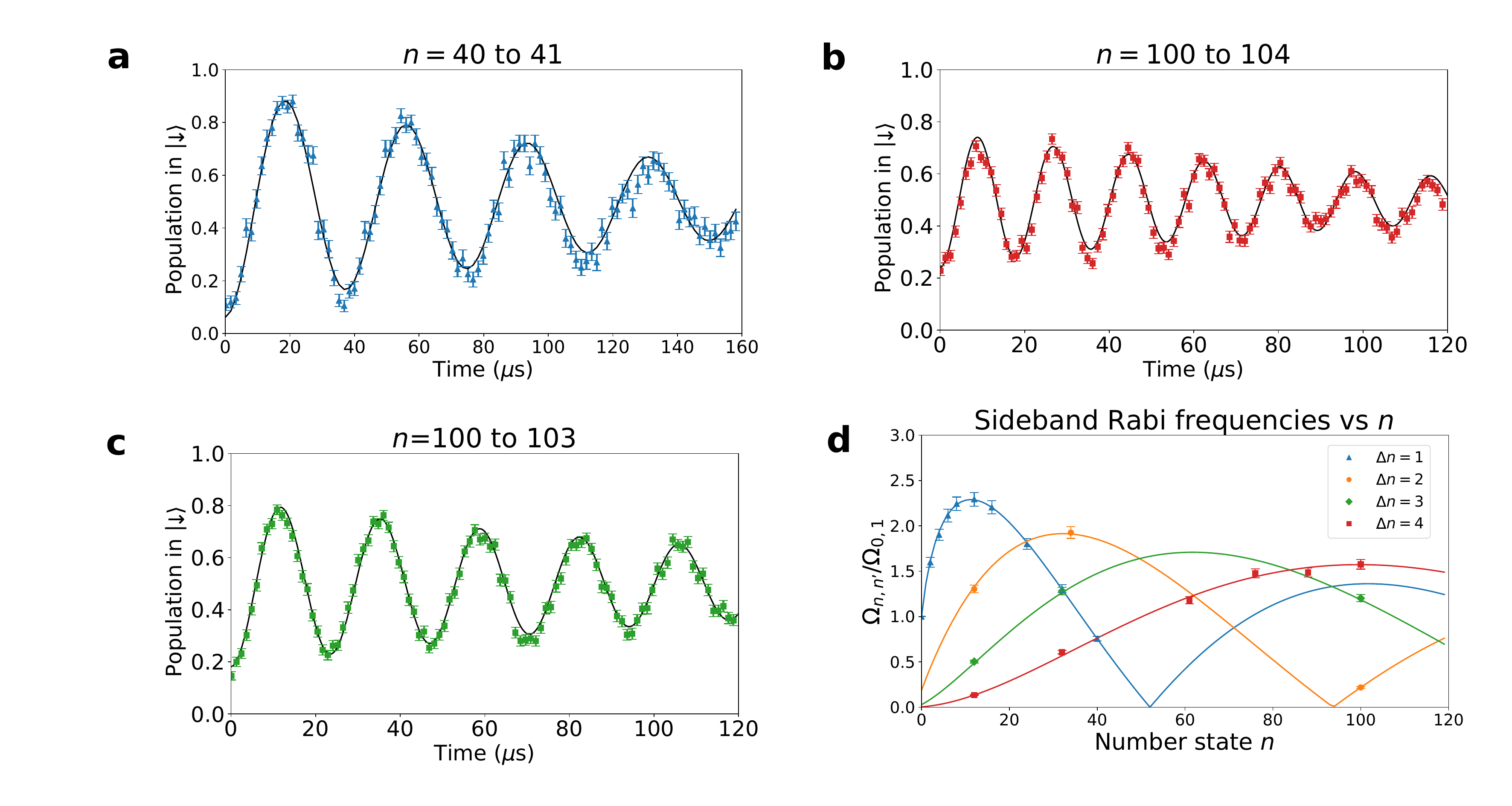}
\caption{\label{Fig:Flop} {\bf Sideband flopping on number states.} {\bf a,} RSB flopping on the 1st-order sideband of an approximate $\lvert \uparrow\rangle\lvert n=40\rangle$ state which is prepared using only 1st-order RSB and BSB pulses. The curve shows the probability of measuring $\lvert \downarrow\rangle$ as a function of pulse duration during 1st RSB flopping to $\lvert \downarrow\rangle\lvert n=41\rangle$. Each point represents an average over 200 experiments. The error bars represent one standard deviation of the mean in all figures. 
The solid black lines in all figures show theory fits to the data with the Rabi frequency, the initial contrast and an exponential decay constant as fit parameters.
{\bf b,} RSB flopping on the 4th-order sideband of an approximate $\lvert \uparrow\rangle\lvert n=100\rangle$ state prepared using 1st to 4th-order RSB and BSB $\pi$-pulses. The curve shows the probability of measuring $\lvert \downarrow\rangle$ as a function of pulse duration during 4th-order RSB flopping to $\lvert \downarrow\rangle\lvert n=104\rangle$. Each point represents an average over 500 experiments. {\bf c,} RSB flopping on the 3rd-order sideband. Preparation as in {\bf b}, then the approximate $\lvert \uparrow\rangle\lvert n=100\rangle$ state is flopped on the 3rd-order sideband to $\lvert \downarrow\rangle\lvert n=103\rangle$ state. Each point represents an average over 500 experiments. {\bf d,} Curves of 1st to 4th-order sideband Rabi frequencies using a fit to all 1st-order data points (blue triangles) to determine the Lamb-Dicke parameter $\eta=0.2632(2)$. Theory curves for higher-order sideband Rabi frequencies are plotted for the same $\eta$. The measured Rabi-frequencies for higher-order sidebands (colored symbols) are consistent with the theoretical Rabi-frequencies (solid colored lines) for all orders. The duration of a $\pi$-pulse from $\lvert n=0 \rangle$ to $\lvert n=1 \rangle$ is approximately 13 $\mu$s. The durations required to produce higher number states can be calculated based on this and the plotted Rabi frequencies for higher number states (see Methods). 
}
\end{figure}
\begin{figure}
\centering
\includegraphics[width = \textwidth]{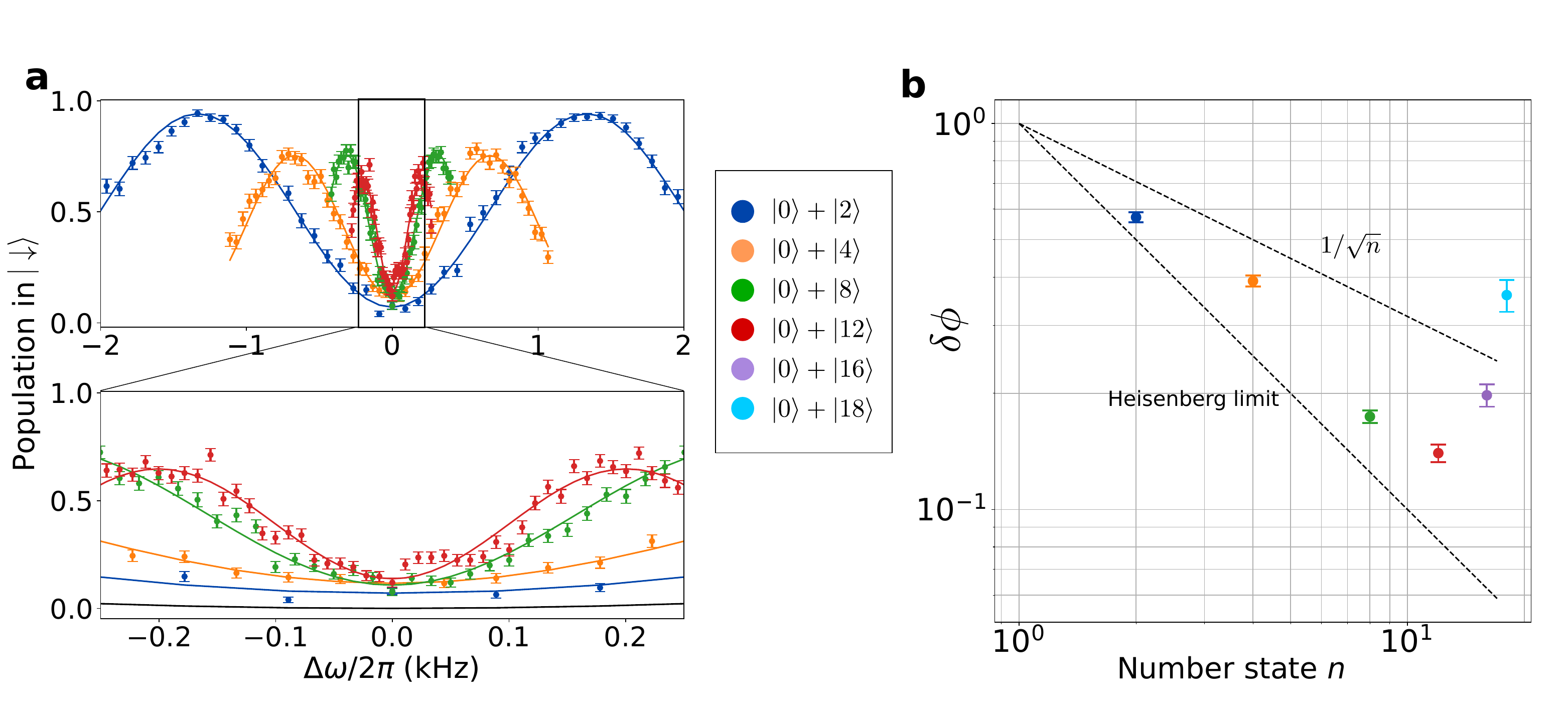}
\caption{\label{Fig:Intf} {\bf Interference and sensitivity of different number-state superpositions.} {\bf a,} Interference fringes for number state interferometers with $n=$ 2, 4, 8 and 12. Each data point is averaged over 250 experiments and uses a wait time of 100 $\mu$s before and after the effective $\pi$-pulse. The fringe spacing is reduced as $1/n$ as expected for Heisenberg scaling. At the same time, the fringe contrast is reduced with higher $n$ due to the larger number of imperfect pulses and the higher susceptibility to mode-frequency changes that are not stable over all 250 experiments for each data point. This reduces the fringe slopes for $n>12$ below the maximal slope reached for $n=12$. Solid lines show theory fits using Eq. \ref{Eq:Prob0} with fringe spacing, contrast and vertical offset as fit parameters. We attribute deviations from expected sinusoidal behavior (for example, near the center of the $\lvert 0 \rangle + \lvert 12 \rangle$ fringe) to changes in the Raman sideband Rabi frequencies by a few percent, temporarily reducing the contrast for few of the points. {\bf b,} Experimentally determined noise-to-signal ratio $\delta \phi$ as defined in Eq.(\ref{Eq:PhaseUncertainty}) as a function of order $n$ (colored dots) together with the theoretical lines for a perfect classical interferometer at $1/\sqrt{n}$ and the $1/n$ Heisenberg limit valid for ideal number-state interferometers.}
\end{figure}
\begin{figure}
\centering
\includegraphics[width = \textwidth]{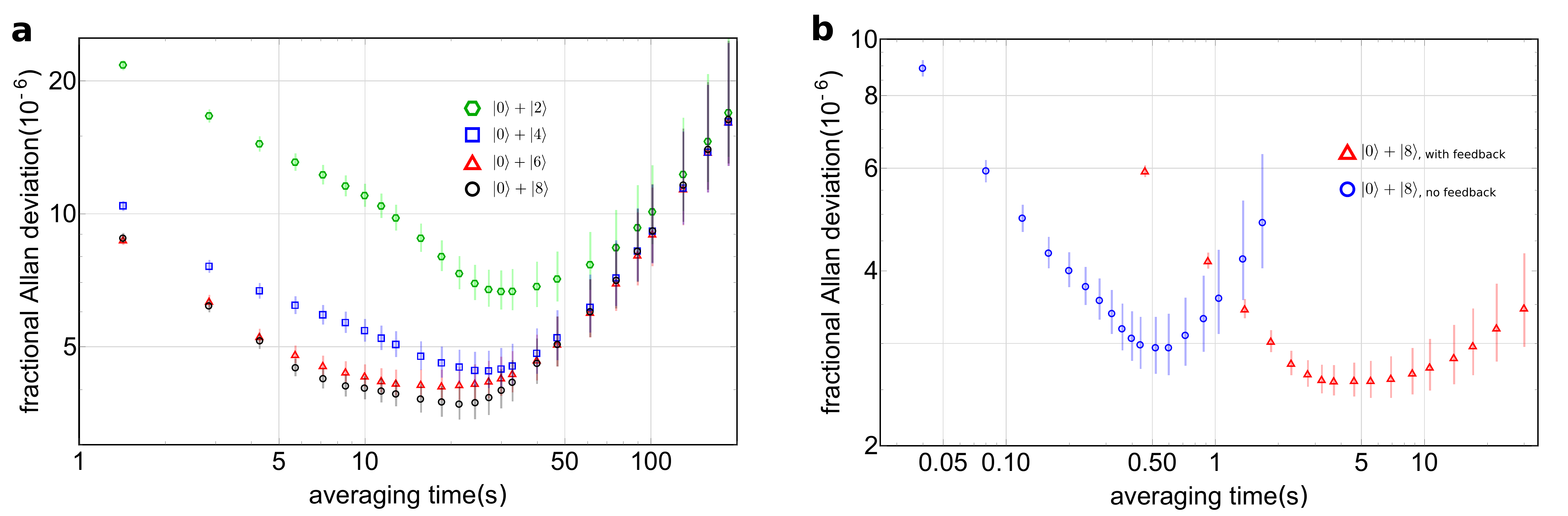}
\caption{\label{Fig:tracking} {\bf Oscillator frequency tracking using number-state interferometers.} \textbf{a,} Interleaved comparison of the Allan standard deviation of tracked fractional trap frequencies vs. averaging time found with $n=2$, $n=4$, $n=6$ and $n=8$ interferometers. The repetition rate of a single run, comprised of long (100 $\mu$s Ramsey time) and short (20 $\mu$s Ramsey time) auto-balance sequences on both sides of the fringe respectively, was approximately 7/s. The $n=8$ interferometer produces the lowest fractional frequency Allan deviation. Trap frequency drifts begin to dominate the Allan deviation at 10's of seconds. \textbf{b,} Fractional mode-frequency uncertainty vs. averaging time for two series of only $n=8$ interferometer runs to maximize measurement duty cycle. We are able to achieve a minimal fractional frequency Allan deviation of 2.6(2) $\times 10^{-6}$ at  approximately 4 seconds of averaging time with tracking activated (red triangles) and an experiment rate of approximately 43/s as defined above. The Allan standard deviation for averaging times up to 1 s without tracking activated is shown by the blue circles. The minimum is reached after 0.5 s with the experiment rate increased to approximately 250/s.}
\end{figure}

%
\textbf{
The use of special quantum states to achieve sensitivities below the limits established by classically behaving states has enjoyed immense success since its inception \cite{Pezz18,Brau18}. In bosonic interferometers, squeezed states\cite{Caves80a,Loud00}, number states\cite{Loud00,Boto00,Haro13} and cat states\cite{Haro13} have been implemented on various platforms and have demonstrated improved measurement precision over interferometers based on coherent states\cite{Schro26,Glaub06}. 
Another metrologically useful state is an equal superposition of two eigenstates with maximally different energies; this state ideally reaches the full interferometric sensitivity allowed by quantum mechanics \cite{Marg98,Giov04,Cave10}. 
By leveraging improvements to our apparatus made primarily to reach higher operation fidelities in quantum information processing, we extend a technique\cite{Meek96} to create number states up to $n=100$ and to generate superpositions of a harmonic oscillator ground state and a number state of the form $\textstyle{\frac{1}{\sqrt{2}}}(\lvert 0\rangle+\lvert n\rangle)$ with $n$ up to 18 in the motion of a single trapped ion. While experimental imperfections prevent us from reaching the ideal Heisenberg limit, we observe enhanced sensitivity to changes in the oscillator frequency that initially increases linearly with $n$, with maximal value at $n=12$ where we observe 3.2(2) dB higher sensitivity compared to an ideal measurement on a coherent state with the same average occupation number. The quantum advantage from using number-state superpositions can be leveraged towards precision measurements on any harmonic oscillator system; here it enables us to track the average fractional frequency of oscillation of a single trapped ion to approximately 2.6 $\times$ 10$^{-6}$ in 5 s. Such measurements should provide improved characterization of imperfections and noise on trapping potentials, which can lead to motional decoherence, a leading source of error in quantum information processing with trapped ions \cite{Ball16,Gaeb16}.\\
\\
}
A large variety of experimental systems, including optical\cite{Sieg00a,Sieg00b}
and microwave\cite{Haro13,Blais04} resonators, micro-mechanical oscillators \cite{Aspe14} and the motion of trapped neutral atoms\cite{Grimm00} and ions\cite{Leib03} can be modeled as harmonic oscillators and controlled in the quantum regime. This opens the possibility of designing and generating advantageous harmonic oscillator quantum states to increase sensitivity or speed when characterizing or controlling the properties of these oscillators. Precisely controlled harmonic oscillators play crucial roles in precision metrology\cite{LIGO13}, fundamental quantum mechanical research\cite{Haro13} and quantum information processing\cite{Ladd10}.\\
\\
The motion of a single atom in a typical trap constitutes a simple mechanical harmonic oscillator. This oscillator can be coupled to internal levels of the atom with electromagnetic fields to cool its motion and transfer motional state information into electronic levels that can be read out by state-selective fluorescence\cite{Leib03}. Ion motional frequencies have been determined previously by resolved-sideband spectroscopy\cite{Leib03} and by excitation with oscillating\cite{Wine83} or pulsed electric fields \cite{Sher11,Alon16}. The sensitivity to excitation from
an oscillating electric field (often called a ``tickle'') can be enhanced by first cooling the motion to the ground state and then detecting small increases in motional energy by exciting red sideband transitions that are energetically forbidden when the ion is in the ground state\cite{Home11}. Here we implement an alternative approach for characterizing motional frequencies that is based on the interference of non-classical number state superpositions\cite{Leib02}. 
These techniques can be especially useful for quantum information processing with trapped ion systems \cite{Cira95}, where qubits are coupled through shared modes of motion. The high fidelity required for fault-tolerant processing makes the stability of the harmonic motion of trapped ions a ubiquitous concern\cite{Ball16,Gaeb16}, particularly as experiments move towards miniaturized traps with smaller ion-to-electrode distances, exposing the ions' motion to larger stray electric fields and increased noise originating from the electrodes\cite{Turch00,Brownnutt15}.\\
\\
Enhanced interferometric sensitivity of superpositions of eigenstates depends on the difference in energy between the two states---in the experiments here where the first state of the superposition is the ground state this energy difference scales linearly with $n$. Here, the harmonic oscillator number operator acts on the number states $\lvert n \rangle$ giving $\hat{a}^\dag \hat{a} \lvert n\rangle =n \lvert n\rangle$, where $\hat{a}$ is the annihilation ladder operator and $n\geq 0$ an integer. We consider a harmonic oscillator with a frequency of oscillation $\omega + \delta\omega(t)$, where $\delta\omega(t)$ is a small, time-dependent correction due to noise and drift relative to the frequency $\omega$ of an ideal reference oscillator, which we will call a local oscillator in keeping with common terminology. We implement a Ramsey-type experiment, where the first effective $\pi/2$-pulse creates the state $\lvert \Psi_n\rangle = \frac{1}{\sqrt{2}}(\lvert 0\rangle + \lvert n\rangle)$, where $n$ denotes the order of the state and interferometer (in the remaining text we will omit normalization). In a frame rotating at the local oscillator frequency, $\lvert 0\rangle$ and $\lvert n\rangle$ will acquire a relative phase between the two states that is proportional to $n$ and the integral of the fluctuations $\delta\omega(t)$ over time $T$. After a duration $T$ the state is
\begin{eqnarray}
\lvert \Psi_n\rangle_T &=& \lvert 0\rangle + e^{i n \phi}\lvert n\rangle ,\nonumber\\
\phi &=& \int_{0}^{T} \delta\omega(t) dt.
\end{eqnarray}
Subsequently, a second effective $\pi/2$-pulse synchronous with the local oscillator recombines the number-state superposition to the ground state if $n\phi= (2 m+1) \pi$ or to $\lvert n\rangle$ if $n\phi= 2 m \pi$ with $m$ an integer. For general $\phi$, the final state (up to a global phase) is $\lvert \Psi_n\rangle_f = \cos (n \phi/2)\lvert n\rangle - i \sin (n \phi/2) \lvert 0\rangle$, with a probability of being in the ground state given by
\begin{equation}\label{Eq:Prob0}
P_0 = 1/2[1-\cos (n \phi)].
\end{equation}
To characterize the harmonic oscillator in interferometric measurements, we want to determine small deviations of $\phi$ around some nominal value with maximal sensitivity. This occurs when the slope $\lvert \partial P_0/\partial \phi\lvert = \lvert n/2~\sin (n \phi)\lvert =n/2$, namely when $n \phi \simeq m \pi/2$, with $m$ an odd integer ($m=\pm 1$ in the tracking experiments described below). We want to minimize
\begin{equation}\label{Eq:PhaseUncertainty}
\delta \phi = \frac{\Delta P_0}{\lvert \partial P_0/\partial \phi\lvert },
\end{equation}
where $\Delta P_0 =\sqrt{\langle P_0^2 \rangle-\langle P_0 \rangle^2}$ is the standard deviation of a population measurement that can discriminate between $\lvert 0\rangle$ and $\lvert n\rangle$. Ideally, the measurement is projection noise limited\cite{Itano93}, which implies $\Delta P_0 = \sqrt{P_0(1-P_0)}$. In this case, $\Delta P_0=1/2$ and $\delta \phi=1/n$, which is the Heisenberg limit and can only be reached with non-classical oscillator states. 
In fact, the state $\lvert 0\rangle + \lvert n\rangle$ satisfies the Margolus-Levitin bound for the maximal rate of evolution \cite{Marg98}, implying that no other combination of states with quantum numbers between 0 and $n$ can produce interference fringes with higher sensitivity to motional frequency changes. The phase uncertainty of classically behaving interferometers, which we define as those that use coherent oscillator states of the same average excitation number $\bar{n}=n/2$ and measurements of the oscillator energy (equivalent to the mean occupation number), will only reduce as $\delta \phi_{\rm class} =\sqrt{1/n}$ (see Methods). \\   
\\
In practice, the effective $\pi/2$-pulses will not have perfect fidelity and there will be added noise above the fundamental projection noise. Such imperfections reduce the contrast $C$ ($0\leq C \leq 1$) from the ideal value $C=1$, which can be incorporated as $P_0 = C/2[1-\cos(n \phi)]$.
In our experiments $C$ decreases as the complexity of preparing superpositions increases with larger $n$. Additionally, since a single experiment only gives us one bit of information (the ion is determined to be in either $\lvert 0\rangle$ or $\lvert n\rangle$), we need to perform multiple experiments to accumulate statistics to determine a phase shift. If the mode-frequency noise is not stable over the time period required to acquire statistics, then the contrast of our interferometer is reduced. This becomes more of a problem with the higher order interferometers, since the susceptibility to mode-frequency noise increases with $n$. This limits the measurable gains in sensitivity to $n\leq 12$ in our specific experimental setting
(see below).\\
\\
Experiments were performed with a single $^9 $Be$ ^+$ ion trapped 40 $\mu$m above a cryogenic ($\simeq$ 4 K) linear surface-electrode trap described elsewhere\cite{Brown11,Wils14}. We use three levels within the electronic $^2 S_{1/2}$ ground-state hyperfine manifold, $\lvert F=1,m_F = -1\rangle = \lvert \uparrow\rangle$, $\lvert F=2,m_F=-2 \rangle = \lvert \downarrow\rangle$, and $\lvert F=2,m_F = 0\rangle = \lvert {\rm aux}\rangle$ to prepare approximate pure number states and number-state superpositions. Here, $F$ is the total angular momentum and $m_F$ is its component along the quantization axis, defined by a 1.43\,
mT static magnetic field. The states $\lvert \downarrow\rangle$ and $\lvert {\rm aux}\rangle$ are shifted from $\lvert \uparrow\rangle$ by $\omega_0 \simeq 2\pi \times - 1.281$ GHz and $\omega_{\rm aux} \simeq 2\pi \times - 1.261$ GHz, respectively (Fig.~1 a). The non-classical motional states are created on the lowest frequency (axial) mode of the three orthogonal harmonic oscillator modes of the ion, with frequency  $\omega \simeq 2\pi \times 7.2$ MHz. The ion is prepared in $\lvert \downarrow\rangle\lvert 0\rangle$ with a fidelity exceeding 0.99 by Doppler laser cooling, followed by ground-state cooling\cite{Leib03, Monr95} (the two transverse modes are only Doppler cooled to $\bar{n} < 1$). Sideband transitions $\lvert \downarrow\rangle\lvert k\rangle \leftrightarrow \lvert \uparrow\rangle\lvert k\pm 1\rangle$ are implemented with stimulated Raman transitions from two photons that are detuned from the $^{2}S_{1/2} \rightarrow$ $^{2}P_{1/2}$ transition ($\lambda \simeq 313$ nm) by approximately +80 GHz\cite{Monr95}. Carrier transitions $\lvert \downarrow\rangle\lvert k\rangle \leftrightarrow \lvert \uparrow\rangle\lvert k\rangle$ and $\lvert \uparrow\rangle\lvert k\rangle \leftrightarrow \lvert {\rm aux}\rangle\lvert k\rangle$ are driven by microwave fields induced by a current through one of the surface trap electrodes.\\
\\
We can distinguish measurements of the $\lvert \uparrow\rangle$ and $\lvert \downarrow\rangle$ states with state-selective fluorescence\cite{Leib03}. When scattering light from a laser beam resonant with the $\lvert \downarrow\rangle \leftrightarrow \lvert P_{3/2}, F=3,m_F=-3\rangle$ cycling transition, 14 to 16 photons are detected on average over 400 $\mu$s with a photomultiplier if the ion is in $\lvert \downarrow\rangle$, while only 0.1 to 0.3 photons (dark counts and stray light) are detected on average if the ion is projected into $\lvert \uparrow\rangle$. We identify outcomes of three or more counts with measurement outcome $\lvert \downarrow\rangle$. Despite deviations from ideal Poisson statistics due to pumping into other hyperfine levels, histograms over many detections on an ion prepared in either $\lvert \downarrow\rangle$ or $\lvert \uparrow\rangle$ reveal
discrimination errors below 0.02. After preparing $\lvert \downarrow \rangle \rightarrow \lvert \uparrow\rangle+\lvert \downarrow\rangle$, we observe an average population in $\lvert \downarrow\rangle$ of 0.489(4) and a variance of 0.249(2) close to the ideal values of 1/2 and 1/4 (All stated uncertainties and error bars in figures represent one standard deviation of the mean).\\
\\
Starting from $\lvert \downarrow\rangle\lvert 0\rangle$, the generation of higher number states is accomplished by first applying a microwave $\pi$-pulse to transform the initial state to $\lvert \uparrow\rangle\lvert 0\rangle$, followed by a series of alternating red ($\lvert \uparrow\rangle\lvert k\rangle \rightarrow \lvert \downarrow\rangle\lvert k+1\rangle$, RSB) and blue ($\lvert \downarrow\rangle\lvert k\rangle \rightarrow \lvert \uparrow\rangle\lvert k+1\rangle$, BSB) sideband $\pi$-pulses at frequencies $\omega_0 - \omega$ and $\omega_0 + \omega$ respectively, so that each pulse flips the spin of the internal state and adds a quantum of motion\cite{Meek96} (Figure \ref{Fig:numberstate} ~a and ~b). In general, we can use sidebands of order $l$ at frequencies $\omega_0 - l\omega$ and $\omega_0 + l\omega$ to add $l$ quanta of motion with a single $\pi$-pulse.  While the Rabi frequencies of higher order sidebands on a mode with Lamb-Dicke parameter $\eta< 1$ are suppressed by $\eta^{l}$ near the ground state, for higher number states, the Rabi frequencies can be much larger than that of the first-order sideband \cite{Leib03} (see Fig.~\ref{Fig:Flop}~{d}).\\
\\
We demonstrate control over the motional state of the ion by preparing it in different (approximately pure) number states and Rabi-flopping on RSBs to determine the contrast, decay, and $n$-dependent Rabi frequency \cite{Leib03} (see Fig.~\ref{Fig:Flop}~a). With the use of only first-order sidebands to create $\lvert \uparrow \rangle \lvert n=40 \rangle$, we are able to achieve RSB flopping ($\lvert \uparrow \rangle \lvert n=40 \rangle \leftrightarrow \lvert \downarrow \rangle \lvert n=41 \rangle $) with greater than 70\,\% contrast. If we make use of up to fourth-order sidebands to create the motional state, we observe approximately 50\,\% contrast when we flop $\lvert n=100\rangle$ on the fourth-order RSB (Fig.~\ref{Fig:Flop}~{b}). To verify that the population participating in the fourth-sideband flopping is in the desired number state, we also flop on the second and third-order sideband (Fig.~\ref{Fig:Flop}~{c}), which have a distinctly different Rabi frequency, and compare the Rabi frequencies of second to fourth order flopping to theory (colored symbols in Fig.~\ref{Fig:Flop}~{d}). For the second sideband, we measured a Rabi frequency (in units of $\Omega_{01}$) of $0.2183 \pm 0.008$ for $n=100$, which agrees with theory within one standard deviation. The calculated Rabi-frequencies in $n=99$ and $n=101$ are more than three standard deviations away from this measured value. Similar comparisons for other values of $n$ further establish confidence that intermediate states are prepared as desired and we can indeed transfer more than $50\,\%$ of the population to $n=100$.\\
\\
The motional superposition $\lvert \uparrow\rangle(\lvert 0\rangle+\lvert 2\rangle)$ is straightforward to prepare by replacing the first RSB pulse by a $\pi/2$-pulse $\lvert \uparrow\rangle\lvert 0\rangle \rightarrow \lvert \uparrow\rangle\lvert 0\rangle+\lvert \downarrow\rangle\lvert 1\rangle$ followed by a BSB $\pi$-pulse, which transforms \linebreak $\lvert \downarrow\rangle\lvert 1\rangle \rightarrow \lvert \uparrow\rangle\lvert 2 \rangle $ while not affecting $\lvert \uparrow\rangle\lvert 0\rangle$ (see Fig.~1 a). To realize the effective $\pi/2$-pulse $\lvert 0 \rangle \rightarrow \lvert 0 \rangle + \lvert n \rangle$ when $n>2$, after the initial RSB $\pi/2$-pulse the $\lvert \uparrow\rangle\lvert 0\rangle$ component is ``shelved'' with a microwave $\pi$-pulse to $\lvert {\rm aux}\rangle\lvert 0\rangle$ that is unaltered by subsequent pairs of BSB and RSB $\pi$-pulses that promote the $\lvert \downarrow\rangle\lvert k\rangle$ component to $\lvert \downarrow\rangle\lvert k+2\rangle$. The preparation is finished by a final microwave $\pi$-pulse $\lvert {\rm aux}\rangle\lvert 0\rangle \rightarrow \lvert \uparrow\rangle\lvert 0\rangle$ and a BSB pulse $\lvert \downarrow\rangle\lvert n-1\rangle+\lvert \uparrow\rangle\lvert 0\rangle \rightarrow \lvert \uparrow\rangle (\lvert 0\rangle+\lvert n\rangle)$ that promotes the $\lvert \downarrow\rangle\lvert n-1\rangle$ component to $\lvert \uparrow\rangle\lvert n\rangle$ while leaving the $\lvert \uparrow\rangle\lvert 0\rangle$ component unaltered.\\
\\
We characterize the enhanced sensitivity of each interferometer by inserting an effective ``spin-echo'' type $\pi$-pulse between wait periods of duration $T$ and purposely induce trap frequency changes $\Delta \omega$ with equal magnitude and opposite sign before and after the echo pulse (Fig.~\ref{Fig:numberstate} c). We use this echo technique because trap frequency fluctuations or drifts that alter the mode frequency by approximately the same amount in both arms are suppressed. For the first wait period this ideally results in an order-dependent phase $\phi=n~ \Delta \omega~ T$. The echo pulse is composed of the following steps: first the pulses of the effective $\pi/2$-pulse are applied in reverse order to ideally give $\lvert \uparrow\rangle (\lvert 0\rangle+e^{i \phi}\lvert n\rangle) \rightarrow \lvert \uparrow\rangle \lvert 0\rangle+e^{i \phi}\lvert \downarrow\rangle\lvert 1\rangle$. Second, a RSB $\pi$-pulse results in $e^{i \phi}\lvert \uparrow\rangle \lvert 0\rangle+\lvert \downarrow\rangle\lvert 1\rangle$, which is then walked up the number-state ladder as described for the first effective $\pi/2$-pulse. Ideally the effective echo $\pi$-pulse accomplishes $\lvert 0\rangle+e^{i \phi}\lvert n\rangle \rightarrow e^{i \phi}\lvert 0\rangle +\lvert n\rangle$. In this way, the induced trap frequency change $-\Delta \omega$ during the second wait period constructively adds to the interferometer phase $e^{i \phi}\lvert 0\rangle+\lvert n\rangle \rightarrow e^{i \phi}\lvert 0\rangle +e^{-i \phi}\lvert n\rangle$, which is transformed to $-i \sin(\phi)\lvert \downarrow\rangle\lvert 1\rangle +\cos(\phi)\lvert \uparrow\rangle)\lvert 0\rangle$ by the final effective $\pi/2$-pulse, so the induced interferometer phase $\phi$ can be read out by measuring the probability to find $\lvert\downarrow\rangle$. \\
\\
We find that as we increase $n$ in the superposition for $T$ = 100 $\mu$s, the phase accumulation increases linearly with $n$ as expected, but, due to accumulated imperfect state preparation steps, the contrast of the interference fringes (and therefore the fringe slope) is reduced (see Fig. \ref{Fig:Intf} a), reducing the signal in Eq.(\ref{Eq:PhaseUncertainty}) of the higher-order interferometers. Given this effect, we observe the highest sensitivity with the $\lvert 0 \rangle + \lvert 12 \rangle$ superposition state, which achieves a 8.5(2) dB (7.1(4) times) improvement over a perfect $\lvert 0\rangle + \lvert 1\rangle$ interferometer (Fig.~\ref{Fig:Intf} b). The $n=12$ interferometer also performs 3.2(2) dB better than an ideal classical interferometer (see Methods). To similarly increase sensitivity with ideal squeezed states\cite{Loud00}, we would require approximately 6 dB of squeezing below the vacuum noise for the ideal interferometer.\\
\\
We can use this enhanced sensitivity to precisely track motional mode frequency changes over time. We perform two Ramsey-type experiments with the phase of the final effective $\pi/2$-pulses equal to $\pm \pi/2$ so that when the pulses are resonant with the mode frequency, the resulting $ \lvert \downarrow \rangle$ population from each Ramsey experiment is ideally 1/2. A difference between the populations for the $+\pi/2$ and $-\pi/2$ cases provides an error signal that we feed back to the local oscillator to follow the fringe pattern as the mode frequency drifts due to changes in stray electric fields and the sources providing the electrode potentials. \\
\\
This procedure is complicated by the fact that the sideband transition frequencies are shifted by the AC Stark effect from the Raman beams. These result in phases beyond those described in Eq. (1) that shift the interferometer fringes. To mitigate this effect, as well as to subtract out non-zero phase accumulation during the creation of the superposition state, we use auto-balanced Ramsey spectroscopy \cite{Sann17} (see Methods). Instead of using two Ramsey experiments that provide the error signal used to feed back to the pulse frequency, we interleave four Ramsey experiments with two different Ramsey times, $T_{short}$ and $T_{long}$ (typically $20 ~\mu$s and $100 ~\mu$s, respectively). The phase between the two $\pi/2$-pulses is adjusted to compensate for systematic phases according to the error signal from the short Ramsey experiments, and the frequency of the local oscillator is adjusted according to that from the long Ramsey experiments. These phase and frequency adjustments are applied equally to both the short- and long-pulse Ramsey experiments. This suppresses all contributions to phase accumulation other than the phase difference accumulated due to the different free-precession times, which is unperturbed by laser beam couplings \cite{Sann17}.\\
\\
When tracking the mode frequency in this way we can record the frequency error versus time for different-$n$ interferometers. We can then determine the overlapping Allan deviation\cite{Howe81} as a function of averaging interval and compare it for the different interferometers (Fig.~\ref{Fig:tracking} a). This data was taken while interleaving experiments with the $\lvert 0 \rangle + \lvert 2 \rangle$, $\lvert 0 \rangle + \lvert 4\rangle$, $\lvert 0\rangle + \lvert 6\rangle$ and $\lvert 0\rangle + \lvert 8\rangle$ interferometers to allow for a direct comparison of their sensitivity under the same noise and drift conditions. For long averaging periods, trap frequency drifts dominate the uncertainty. As expected, this gives the same asymptotic long-time slope of the Allan deviation for all interferometers. Importantly, the increased sensitivity of higher-$n$ interferometers reduces the time interval required to average down to a certain level for $n$ up to 8. For the $n=8$ interferometer, we observe the minimum Allan standard deviation at approximately 23 s of averaging in this interleaved comparison. By running only the $n=8$ interferometer sequences, we increase the measurement duty cycle which accelerates the rate with which the Allan deviation approaches its minimum. Under these conditions, the minimal fractional frequency Allan deviation of 2.6(2)$\times 10^{-6}$ ($\sim$ 19 Hz at 7.2 MHz) is reached at approximately 4 seconds of averaging (red triangles in
Fig.~\ref{Fig:tracking} b). To further increase the measurement rate, we record the population differences determined in all four Ramsey experiments comprising the auto-balance sequence without feeding back on the local oscillator frequency. This eliminates the latency due to computer control of the frequency tracking. As long as the populations of four Ramsey experiments uniquely determine the frequency change, we can run a series of $n=8$ interferometer experiments, each taking 4 ms, without feedback to shorten the time to reach the minimum of 2.9(4) $\times 10^{-6}$ to approximately 0.5 s before the uncompensated mode frequency drift produces an increasing Allan deviation (Fig.~\ref{Fig:tracking} b, blue circles). While the minimum value of the Allan deviation is not lower when taking data in this fashion as compared to when tracking the drift, this experiment gives an idea of how quickly we could average down to the level of a few parts in 10$^{-6}$ if tracking latency is minimized.\\
\\
In conclusion, we have demonstrated preparation of approximate number states for the harmonic motion of a trapped ion up to $n = 100$ and characterized quantum-enhanced sensitivity of number state interferometers up to $n=18$. 
We used this sensitivity to measure the mode frequency with a minimum fractional frequency uncertainty of $2.6(2) \times 10^{-6}$. The quantum advantages we demonstrate were reduced by imperfections in state preparation and detection, as well as uncontrolled mode-frequency changes, most likely caused by time-varying stray fields and technical noise on the potentials applied to trap electrodes (see Methods). In a natural extension of the work presented here, it should be possible to observe such mode-frequency noise during free precession by refocusing with one or more effective $\pi$-pulses. This would allow us to filter the response of the ion to certain spectral components of the motional mode-frequency noise, providing a quantum lock-in analyzer in analogy to characterizations of magnetic field noise with a trapped ion\cite{Kotl11}. By using number-state superpositions, we can transfer the quantum advantage demonstrated in the experiment reported here to achieve ``quantum gain'' in such lock-in measurements. \\
\\
More generally, we expect that the techniques demonstrated here, as well as an alternative approach in Ref.{\renewcommand\citemid{}\cite[\hspace*{0.8pt}]{Wolf18}}, can be applied to characterize other harmonic oscillators in the quantum regime with increased precision and on time scales that were previously inaccessible. Such capabilities could support quantum metrology and improve the prospects of fault-tolerant quantum information processing, where some of the most advanced experimental platforms are limited by harmonic oscillator coherence.
\newpage
%
\bibliographystyle{naturemag}
\bibliography{refs}

\begin{thebibliography}{10}
\expandafter\ifx\csname url\endcsname\relax
  \def\url#1{\texttt{#1}}\fi
\expandafter\ifx\csname urlprefix\endcsname\relax\def\urlprefix{URL }\fi
\providecommand{\bibinfo}[2]{#2}
\providecommand{\eprint}[2][]{\url{#2}}

\bibitem{Pezz18}
\bibinfo{author}{Pezz\`e, L.}, \bibinfo{author}{Smerzi, A.},
  \bibinfo{author}{Oberthaler, M.~K.}, \bibinfo{author}{Schmied, R.} \&
  \bibinfo{author}{Treutlein, P.}
\newblock \bibinfo{title}{Quantum metrology with nonclassical states of atomic
  ensembles}.
\newblock \emph{\bibinfo{journal}{Rev. Mod. Phys.}}
  \textbf{\bibinfo{volume}{90}}, \bibinfo{pages}{035005}
  (\bibinfo{year}{2018}).

\bibitem{Brau18}
\bibinfo{author}{Braun, D.} \emph{et~al.}
\newblock \bibinfo{title}{Quantum-enhanced measurements without entanglement}.
\newblock \emph{\bibinfo{journal}{Rev. Mod. Phys.}}
  \textbf{\bibinfo{volume}{90}}, \bibinfo{pages}{035006}
  (\bibinfo{year}{2018}).

\bibitem{Caves80a}
\bibinfo{author}{Caves, C.~M.}, \bibinfo{author}{Thorne, K.~S.},
  \bibinfo{author}{Drever, R. W.~P.}, \bibinfo{author}{Sandberg, V.~D.} \&
  \bibinfo{author}{Zimmermann, M.}
\newblock \bibinfo{title}{On the measurement of a weak classical force coupled
  to a quantum-mechanical oscillator. {I. I}ssues of principle}.
\newblock \emph{\bibinfo{journal}{Rev. Mod. Phys.}}
  \textbf{\bibinfo{volume}{52}}, \bibinfo{pages}{341--392}
  (\bibinfo{year}{1980}).

\bibitem{Loud00}
\bibinfo{author}{Loudon, R.}
\newblock \emph{\bibinfo{title}{The Quantum Theory of Light}}
  (\bibinfo{publisher}{Oxford University Press}, \bibinfo{address}{Great
  Clarendon Street, Oxford, OX2 6DP, UK}, \bibinfo{year}{2000}).

\bibitem{Boto00}
\bibinfo{author}{Boto, A.~N.} \emph{et~al.}
\newblock \bibinfo{title}{Quantum interferometric optical lithography:
  Exploiting entanglement to beat the diffraction limit}.
\newblock \emph{\bibinfo{journal}{Phys. Rev. Lett.}}
  \textbf{\bibinfo{volume}{85}}, \bibinfo{pages}{2733--2736}
  (\bibinfo{year}{2000}).

\bibitem{Haro13}
\bibinfo{author}{Haroche, S.} \& \bibinfo{author}{Raymond, J.-M.}
\newblock \emph{\bibinfo{title}{Exploring the Quantum}}
  (\bibinfo{publisher}{Oxford University Press}, \bibinfo{address}{Great
  Clarendon Street, Oxford, OX2 6DP, UK}, \bibinfo{year}{2006}).

\bibitem{Schro26}
\bibinfo{author}{Schr{\"o}dinger, E.}
\newblock \bibinfo{title}{Der stetige {{\"U}bergang} von der {M}ikro- zur
  {M}akromechanik}.
\newblock \emph{\bibinfo{journal}{Die Naturwissenschaften}}
  \textbf{\bibinfo{volume}{14}}, \bibinfo{pages}{664--666}
  (\bibinfo{year}{1926}).

\bibitem{Glaub06}
\bibinfo{author}{Glauber, R.~J.}
\newblock \bibinfo{title}{Nobel {L}ecture: {One Hundred Years of Light
  Quanta}}.
\newblock \emph{\bibinfo{journal}{Rev. Mod. Phys.}}
  \textbf{\bibinfo{volume}{78}}, \bibinfo{pages}{1267--1278}
  (\bibinfo{year}{2006}).

\bibitem{Marg98}
\bibinfo{author}{Margolus, N.} \& \bibinfo{author}{Levitin, L.~B.}
\newblock \bibinfo{title}{The maximum speed of dynamical evolution}.
\newblock \emph{\bibinfo{journal}{Physica D: Nonlinear Phenomena}}
  \textbf{\bibinfo{volume}{120}}, \bibinfo{pages}{188 -- 195}
  (\bibinfo{year}{1998}).

\bibitem{Giov04}
\bibinfo{author}{Giovannetti, V.}, \bibinfo{author}{Lloyd, S.} \&
  \bibinfo{author}{Maccone, L.}
\newblock \bibinfo{title}{Quantum enhanced measurement: Beating the standard
  quantum limit.}
\newblock \emph{\bibinfo{journal}{Science}} \textbf{\bibinfo{volume}{306}}
  (\bibinfo{year}{2004}).

\bibitem{Cave10}
\bibinfo{author}{Caves, C.~M.} \& \bibinfo{author}{Shaji, A.}
\newblock \bibinfo{title}{Quantum-circuit guide to optical and atomic
  interferometry}.
\newblock \emph{\bibinfo{journal}{Opt. Comm.}} \textbf{\bibinfo{volume}{283}},
  \bibinfo{pages}{695--712} (\bibinfo{year}{2010}).

\bibitem{Meek96}
\bibinfo{author}{Meekhof, D.~M.}, \bibinfo{author}{Monroe, C.},
  \bibinfo{author}{King, B.~E.}, \bibinfo{author}{Itano, W.~M.} \&
  \bibinfo{author}{Wineland, D.~J.}
\newblock \bibinfo{title}{Generation of nonclassical motional states of a
  trapped atom}.
\newblock \emph{\bibinfo{journal}{Phys. Rev. Lett.}}
  \textbf{\bibinfo{volume}{77}}, \bibinfo{pages}{2346--2346}
  (\bibinfo{year}{1996}).

\bibitem{Ball16}
\bibinfo{author}{Ballance, C.~J.}, \bibinfo{author}{Harty, T.~P.},
  \bibinfo{author}{Linke, N.~M.}, \bibinfo{author}{Sepiol, M.~A.} \&
  \bibinfo{author}{Lucas, D.~M.}
\newblock \bibinfo{title}{High-fidelity quantum logic gates using trapped-ion
  hyperfine qubits}.
\newblock \emph{\bibinfo{journal}{Phys. Rev. Lett.}}
  \textbf{\bibinfo{volume}{117}}, \bibinfo{pages}{060504}
  (\bibinfo{year}{2016}).

\bibitem{Gaeb16}
\bibinfo{author}{Gaebler, J.~P.} \emph{et~al.}
\newblock \bibinfo{title}{High-fidelity universal gate set for
  ${^{9}\mathrm{Be}}^{+}$ ion qubits}.
\newblock \emph{\bibinfo{journal}{Phys. Rev. Lett.}}
  \textbf{\bibinfo{volume}{117}}, \bibinfo{pages}{060505}
  (\bibinfo{year}{2016}).

\bibitem{Sieg00a}
\bibinfo{author}{Siegman, A.~E.}
\newblock \bibinfo{title}{Laser beams and resonators: {T}he 1960s}.
\newblock \emph{\bibinfo{journal}{IEEE Journal of Selected Topics in Quantum
  Electronics}} \textbf{\bibinfo{volume}{6}}, \bibinfo{pages}{1380--1388}
  (\bibinfo{year}{2000}).

\bibitem{Sieg00b}
\bibinfo{author}{Siegman, A.~E.}
\newblock \bibinfo{title}{Laser beams and resonators: Beyond the 1960s}.
\newblock \emph{\bibinfo{journal}{IEEE Journal of Selected Topics in Quantum
  Electronics}} \textbf{\bibinfo{volume}{6}}, \bibinfo{pages}{1389--1399}
  (\bibinfo{year}{2000}).

\bibitem{Blais04}
\bibinfo{author}{Blais, A.}, \bibinfo{author}{Huang, R.-S.},
  \bibinfo{author}{Wallraff, A.}, \bibinfo{author}{Girvin, S.~M.} \&
  \bibinfo{author}{Schoelkopf, R.~J.}
\newblock \bibinfo{title}{Cavity quantum electrodynamics for superconducting
  electrical circuits: An architecture for quantum computation}.
\newblock \emph{\bibinfo{journal}{Phys. Rev. A}} \textbf{\bibinfo{volume}{69}},
  \bibinfo{pages}{062320} (\bibinfo{year}{2004}).

\bibitem{Aspe14}
\bibinfo{author}{Aspelmeyer, M.}, \bibinfo{author}{Kippenberg, T.} \&
  \bibinfo{author}{Marquardt, F.}
\newblock \bibinfo{title}{Cavity optomechanics}.
\newblock \emph{\bibinfo{journal}{Rev. Mod. Phys.}}
  \textbf{\bibinfo{volume}{86}}, \bibinfo{pages}{1391--1452}
  (\bibinfo{year}{2014}).

\bibitem{Grimm00}
\bibinfo{author}{Grimm, R.}, \bibinfo{author}{{Weidem{\"u}ller}, M.} \&
  \bibinfo{author}{Ovchinnikov, Y.~B.}
\newblock \bibinfo{title}{Optical dipole traps for neutral atoms}.
\newblock vol.~\bibinfo{volume}{42} of \emph{\bibinfo{series}{Advances In
  Atomic, Molecular, and Optical Physics}}, \bibinfo{pages}{95 -- 170}
  (\bibinfo{publisher}{Academic Press}, \bibinfo{year}{2000}).

\bibitem{Leib03}
\bibinfo{author}{Leibfried, D.}, \bibinfo{author}{Blatt, R.},
  \bibinfo{author}{Monroe, C.} \& \bibinfo{author}{Wineland, D.~J.}
\newblock \bibinfo{title}{Quantum dynamics of single trapped ions}.
\newblock \emph{\bibinfo{journal}{Rev. Mod. Phys.}}
  \textbf{\bibinfo{volume}{75}}, \bibinfo{pages}{281--324}
  (\bibinfo{year}{2003}).

\bibitem{LIGO13}
\bibinfo{author}{Aasi, J.} \emph{et~al.}
\newblock \bibinfo{title}{Enhanced sensitivity of the {LIGO} gravitational wave
  detector by using squeezed states of light}.
\newblock \emph{\bibinfo{journal}{Nature Photonics}}
  \textbf{\bibinfo{volume}{7}}, \bibinfo{pages}{613--619}
  (\bibinfo{year}{2013}).

\bibitem{Ladd10}
\bibinfo{author}{Ladd, T.~D.} \emph{et~al.}
\newblock \bibinfo{title}{Quantum computers}.
\newblock \emph{\bibinfo{journal}{Nature}} \textbf{\bibinfo{volume}{464}},
  \bibinfo{pages}{45--53} (\bibinfo{year}{2010}).

\bibitem{Wine83}
\bibinfo{author}{Wineland, D.~J.}, \bibinfo{author}{Bollinger, J.~J.} \&
  \bibinfo{author}{Itano, W.~M.}
\newblock \bibinfo{title}{Laser fluorescence mass spectroscopy}.
\newblock \emph{\bibinfo{journal}{Phys. Rev. Lett.}}
  \textbf{\bibinfo{volume}{50}}, \bibinfo{pages}{628–631}
  (\bibinfo{year}{1983}).

\bibitem{Sher11}
\bibinfo{author}{Sheridan, K.} \& \bibinfo{author}{Keller, M.}
\newblock \bibinfo{title}{Weighing of trapped ion crystals and its
  applications}.
\newblock \emph{\bibinfo{journal}{New Journal of Physics}}
  \textbf{\bibinfo{volume}{13}}, \bibinfo{pages}{123002}
  (\bibinfo{year}{2011}).

\bibitem{Alon16}
\bibinfo{author}{Alonso, J.} \emph{et~al.}
\newblock \bibinfo{title}{Generation of large coherent states by bang--bang
  control of a trapped-ion oscillator}.
\newblock \emph{\bibinfo{journal}{Nature Communications}}
  \textbf{\bibinfo{volume}{7}}, \bibinfo{pages}{11243} (\bibinfo{year}{2016}).

\bibitem{Home11}
\bibinfo{author}{Home, J.~P.}, \bibinfo{author}{Hanneke, D.},
  \bibinfo{author}{Jost, J.~D.}, \bibinfo{author}{Leibfried, D.} \&
  \bibinfo{author}{Wineland, D.~J.}
\newblock \bibinfo{title}{Normal modes of trapped ions in the presence of
  anharmonic trap potentials}.
\newblock \emph{\bibinfo{journal}{New Journal of Physics}}
  \textbf{\bibinfo{volume}{13}}, \bibinfo{pages}{073026}
  (\bibinfo{year}{2011}).

\bibitem{Leib02}
\bibinfo{author}{Leibfried, D.} \emph{et~al.}
\newblock \bibinfo{title}{Trapped-ion quantum simulator: Experimental
  application to nonlinear interferometers}.
\newblock \emph{\bibinfo{journal}{Phys. Rev. Lett.}}
  \textbf{\bibinfo{volume}{89}}, \bibinfo{pages}{247901}
  (\bibinfo{year}{2002}).

\bibitem{Cira95}
\bibinfo{author}{Cirac, J.~I.} \& \bibinfo{author}{Zoller, P.}
\newblock \bibinfo{title}{Quantum computations with cold trapped ions}.
\newblock \emph{\bibinfo{journal}{Phys. Rev. Lett.}}
  \textbf{\bibinfo{volume}{74}}, \bibinfo{pages}{4091--4094}
  (\bibinfo{year}{1995}).

\bibitem{Turch00}
\bibinfo{author}{Turchette, Q.~A.} \emph{et~al.}
\newblock \bibinfo{title}{Heating of trapped ions from the quantum ground
  state}.
\newblock \emph{\bibinfo{journal}{Phys. Rev. A}} \textbf{\bibinfo{volume}{61}},
  \bibinfo{pages}{063418} (\bibinfo{year}{2000}).

\bibitem{Brownnutt15}
\bibinfo{author}{Brownnutt, M.}, \bibinfo{author}{Kumph, M.},
  \bibinfo{author}{Rabl, P.} \& \bibinfo{author}{Blatt, R.}
\newblock \bibinfo{title}{Ion-trap measurements of electric-field noise near
  surfaces}.
\newblock \emph{\bibinfo{journal}{Rev. Mod. Phys.}}
  \textbf{\bibinfo{volume}{87}}, \bibinfo{pages}{1419--1482}
  (\bibinfo{year}{2015}).

\bibitem{Itano93}
\bibinfo{author}{Itano, W.~M.} \emph{et~al.}
\newblock \bibinfo{title}{Quantum projection noise: Population fluctuations in
  two-level systems}.
\newblock \emph{\bibinfo{journal}{Phys. Rev. A}} \textbf{\bibinfo{volume}{47}},
  \bibinfo{pages}{3554--3570} (\bibinfo{year}{1993}).

\bibitem{Brown11}
\bibinfo{author}{Brown, K.~R.} \emph{et~al.}
\newblock \bibinfo{title}{Coupled quantized mechanical oscillators}.
\newblock \emph{\bibinfo{journal}{Nature}} \textbf{\bibinfo{volume}{471}},
  \bibinfo{pages}{196--199} (\bibinfo{year}{2011}).

\bibitem{Wils14}
\bibinfo{author}{Wilson, A.~C.} \emph{et~al.}
\newblock \bibinfo{title}{Tunable spin--spin interactions and entanglement of
  ions in separate potential wells}.
\newblock \emph{\bibinfo{journal}{Nature}} \textbf{\bibinfo{volume}{512}},
  \bibinfo{pages}{57--60} (\bibinfo{year}{2014}).

\bibitem{Monr95}
\bibinfo{author}{Monroe, C.} \emph{et~al.}
\newblock \bibinfo{title}{Resolved-{S}ideband {R}aman {C}ooling of a {B}ound
  {A}tom to the {3D Zero-Point Energy}}.
\newblock \emph{\bibinfo{journal}{Phys. Rev. Lett.}}
  \textbf{\bibinfo{volume}{75}}, \bibinfo{pages}{4011--4014}
  (\bibinfo{year}{1995}).

\bibitem{Sann17}
\bibinfo{author}{Sanner, C.}, \bibinfo{author}{Huntemann, N.},
  \bibinfo{author}{Lange, R.}, \bibinfo{author}{Tamm, C.} \&
  \bibinfo{author}{Peik, E.}
\newblock \bibinfo{title}{Autobalanced {R}amsey spectroscopy}.
\newblock \emph{\bibinfo{journal}{Phys. Rev. Lett.}}
  \textbf{\bibinfo{volume}{120}} (\bibinfo{year}{2017}).

\bibitem{Howe81}
\bibinfo{author}{Howe, D.~A.}, \bibinfo{author}{W., A.~D.} \&
  \bibinfo{author}{Barnes, J.~A.}
\newblock \bibinfo{title}{Properties of signal sources and measurement
  methods}.
\newblock \emph{\bibinfo{journal}{Proc. 1981 Freq. Cont. Symp.}}
  \bibinfo{pages}{1--47} (\bibinfo{year}{1981}).

\bibitem{Kotl11}
\bibinfo{author}{Kotler, S.}, \bibinfo{author}{Akerman, N.},
  \bibinfo{author}{Glickman, Y.}, \bibinfo{author}{Keselman, A.} \&
  \bibinfo{author}{Ozeri, R.}
\newblock \bibinfo{title}{Single ion quantum lock-in amplifier}.
\newblock \emph{\bibinfo{journal}{Nature}} \textbf{\bibinfo{volume}{473}},
  \bibinfo{pages}{61--65} (\bibinfo{year}{2011}).

\bibitem{Wolf18}
\bibinfo{author}{Wolf, F.} \emph{et~al.}
\newblock \bibinfo{title}{Motional {F}ock states for quantum-enhanced amplitude
  and phase measurements with trapped ions.} \bibinfo{pages}{Preprint at
  https://arxiv.org/abs/1807.01875} (\bibinfo{year}{2018}).

\end{thebibliography}
\section*{Acknowledgements} 
We thank David Allcock, Daniel Slichter and Raghavendra Srinivas for helpful discussions and assistance with the experimental setup and Holly F. Leopardi and Hannah Knaack for useful comments on the manuscript. This work was supported by IARPA, ARO, ONR and the NIST Quantum Information Program. K.C.M. acknowledges support by an ARO QuaCGR fellowship through grant W911NF-14-1-0079. J.K. acknowledges support by the Alexander von Humboldt foundation.
\section*{Author Contributions} 
K.C.M. and D.L. conceived the experiments, carried out the measurements, analyzed the data and wrote the main part of the manuscript. K.C.M., J.K., S.C.B. and A.C.W. built and maintained the experimental setup. D.J.W., A.C.W. and D.L. developed parts of the experimental setup and supervised the work. All authors discussed the results and contributed to the manuscript. 
\section*{Author Information} 
The authors declare that they have no competing financial interests. Correspondence and requests for materials should be addressed to K.C.M. email: katherine.mccormick@nist.gov.
%
%
\section*{Methods}
\subsection*{Precision limit of ``classical'' measurements on ``classical'' states}
 There is fairly general agreement in the literature that any ``classical'' limit to measurement precision on the frequency of an harmonic oscillator will scale as proportional to $\sqrt{\bar{n}}$, where $\bar{n}$ is the average occupation number\cite{Brau18}. Despite this, there is no general agreement on the pre-factor to that scaling, which is necessary to fully define a ``standard quantum limit."\cite{Brau18} The limit we establish assumes an observer with classical resources (defined below) that are perfectly implemented.\\
 \\
For comparisons of our experiments to a well-defined classical reference experiment, we extend the notions of classical experiments with light fields given by Roy Glauber\cite{Glaub06} to a general harmonic oscillator. For light fields, Glauber restricted classical sources to coherent light. In this spirit, we broaden the possible physical implementations to any system that can be described as a harmonic oscillator, but limit the admissible operations to coherent displacements. Harmonic oscillator quantum observables are expressed with ladder operators $\hat{a}$ and $\hat{a}^\dag$ and a coherent state is an eigenstate of the annihilation operator with $\hat{a} \lvert \alpha\rangle = \alpha \lvert \alpha\rangle$. The oscillator's average number of quanta is then given by the expectation value $\bar{n}$ of the number operator $\bar{n}=\langle \alpha \lvert \hat{a}^\dag \hat{a}\lvert \alpha\rangle = \lvert \alpha\lvert ^2$. Glauber restricted classical measurements to measuring intensities for light fields (for example the intensities arising on a screen due to the interference of the two light fields from a double slit arrangement). We generalize light field intensity to number expectation values $\bar{n}$ as the permitted classical measurements in a harmonic oscillator. The attainable signal-to-noise of such measurements will be limited by shot-noise, given by the standard deviation of a Poisson distribution with mean $\bar{n}$, $\Delta P_{\bar{n}}=\sqrt{\bar{n}}$, for an ideal measurement on a coherent state with unit quantum efficiency and no excess noise. To compare to number-state superpositions of the form $\lvert \Psi_n\rangle=\lvert 0\rangle+\lvert n\rangle$, we require that the classical interferometer uses no more energy than the competing number-state interferometer, $\bar{n} = \langle \Psi_n\lvert \hat{a}^\dag \hat{a}\lvert \Psi_n\rangle \leq n$. This definition of equal resources is somewhat arbitrary, for example one could also argue for the same maximal energy. However, the coherent states have no well defined maximal energy and it is always possible to rescale from our definition to another definition of equal resources. The scaling factor would likely be of order unity in most cases and irrespective of its value, the ideal non-classical interferometer will eventually outperform the classical counterpart due to its more favorable scaling in $\bar{n}$.\\
\\
With the definitions above, we can devise an interferometer experiment that has the salient features of a Ramsey experiment, but is based on classical states and measurements. A Ramsey experiment consists of two excitations with known relative phase, separated by ``free precession'' for duration $T$. An example of a classical-like Ramsey experiment could be to send an rf-pulse from a reference oscillator with known frequency and phase into a near-resonant circuit. The pulse will ring up the circuit which then evolves freely for $T$. A second pulse is the sent to the circuit and depending on its phase relative to the first pulse and the phase the excitation has picked up in the resonant circuit during $T$, the two pulses interfere constructively to further build up the field in the circuit or interfere destructively, diminishing the circuit excitation. For a general harmonic oscillator, this can conveniently be described in phase-space in a frame oscillating at the frequency of the reference oscillator. Starting in the ground state, the first excitation creates a coherent state $\lvert \alpha_1\rangle$, where we can choose phase space coordinates that make $\alpha_1 \geq 0$ real without losing generality. During a free-precession time $T$ the state picks up a phase $\phi_T$ which transforms it to $\lvert \alpha_1\rangle_T=\lvert \alpha_1 \exp(i \phi_T)\rangle$. In analogy to a Ramsey experiment, the second excitation is chosen to have the same magnitude with phase $\phi$ relative to the first excitation, $\alpha_2 = \alpha_1 \exp(i \phi)$. This transforms the state (up to a global phase that is of no consequence to this experiment) into $\lvert \alpha \rangle = \lvert \alpha_1 (1 + \exp(i (\phi_T - \phi)))\rangle$. Following Glauber's definition of classical measurements, we measure the average occupation $\bar{n}$:

\begin{equation}
\bar{n} = 4\alpha_1^2(\frac{1}{2} + \frac{1}{2}\cos(\phi-\phi_T))
\end{equation}

For a fair comparison to a number-state superposition $\lvert0\rangle + \lvert n \rangle$, we would like to restrict our resources to $\bar{n} \leq n$ for all possible $\phi$ and $\phi_T$, so we choose the maximum value $4\alpha_1^2 = n$. Under this restriction, we recover a classical version of the Ramsey fringes described by Eq. (\ref{Eq:Prob0}), where we measure $\bar{n} = 0$ when $\phi - \phi_T = (2m+1)\pi$ and $\bar{n} = n$ when $\phi - \phi_T = 2m\pi$, with $m$ integer. We want to minimize the noise-to-signal ratio Eq.(\ref{Eq:PhaseUncertainty}) restricted to coherent excitations and a small free-precession phase $\lvert \phi_T\lvert \simeq 0 \ll\pi$. The noise-to-signal ratio to minimize is
\begin{equation}\label{Eq:ClassNtoS}
\delta \phi_c = \frac{\Delta P_{\bar{n}}}{\lvert \partial P_{\bar{n}}/\partial \phi_T \lvert } \Bigg\lvert_{\phi_T=0} =\frac{\sqrt{n(1/2+1/2\cos(\phi))}}{n/2\sin(\phi)}  =\sqrt{\frac{1}{n}}\sqrt{\frac{1+1/\cos^2(\phi)}{2}}.
\end{equation}
This occurs when $\phi=\pi$, which describes two equal and opposite displacements that put the harmonic oscillator back to the vacuum state if the free-precession phase $\phi_T=0$. Noise and signal both vanish for this interferometer in such a way that their ratio stays finite at $\delta \phi_c=1/\sqrt{n}$.\\
\\
This is the best one could possibly do in the case of no excess noise in the system. In practice, when some added noise would be present, we would choose a value of $\phi$ that maximizes the signal, as we did for the case of the number state interferometer. In this case, as was also true for the number state interferometer, the point of maximal signal occurs for $\phi=\pi/2$, so the practical noise-to-signal-ratio yields:

\begin{equation}\label{Eq:NtoSres2}
\delta \phi_p=\frac{\sqrt{n(1/2+1/2\cos(\phi))}}{n/2\sin(\phi)}\Bigg\lvert_{\phi=\pi/2} = \sqrt{\frac{2}{n}}.
\end{equation}

This interferometer has no practical drawbacks, optimally suppresses additional noise and leads to an increase of the noise-to-signal over the optimal value by $\sqrt{2}$. This implies that the advantage of a number-state interferometer may increase by $\sqrt{2}$ over the ideal limit in a realistic setting, where excess noise (such as current noise or dark counts of a detector) is almost inevitable. This also is arguably a more direct comparison to the $\lvert 0 \rangle + \lvert n \rangle$ interferometer experiment, where we choose the same relative phase between the two pulses of $\phi=\pi/2$ to maximize the signal. Despite the case to be made for using the more practical limit in our comparisons, we strictly use the more stringent ideal classical interferometer in all our sensitivity comparisons of the number-state interferometers in the main text.
\subsection*{Experimental details}

\subsubsection*{Time scales of experiments}
For all of the experiments described in this paper, the state preparation consists of Doppler cooling ($\sim 120~\mu$s) followed by ground state cooling ($\sim 110~\mu$s) and a microwave carrier $\pi$-pulse $\lvert\downarrow\rangle \rightarrow \lvert\uparrow\rangle$ ($\sim 5~\mu$s). At the end of each experiment, a 400 $\mu$s detection pulse on the cycling transition is applied. The Raman red sideband  $\lvert \uparrow \rangle \lvert 0 \rangle \rightarrow \lvert \downarrow \rangle \lvert 1 \rangle$ $\pi$-time, which sets the Rabi-frequency $\Omega_{0,1}$, is approximately 13 $\mu$s, so the duration of individual experiments can be calculated by using $\Omega_{0,1}$ as a ``base" unit to calculate the sideband $\pi$-times for higher $n$ according to\cite{Leib03}:
\begin{equation}
    \Omega_{n,n+s}/\Omega_{0,1}=\eta^{|s|-1}\sqrt{\frac{n_<!}{n_>!}}L_{n_<}^{|s|}(\eta^2)
\end{equation}
where $L_n^{|s|}$ is the generalized Laguerre polynomial and $n_>$ ($n_<$) is the greater (lesser) of $n+s$ and $n$ (see also Fig. \ref{Fig:Flop}d). We can then sum the durations of the individual sideband pulses (and, in the case of the superposition states, the durations of the $\sim 5~\mu$s microwave $\lvert \uparrow \rangle \rightarrow \lvert aux \rangle$ $\pi$-pulses) used. Table \ref{Tab:exptimes} lists durations of the sequences used to generate pure number states and number state superpositions. It is possible to substantially decrease these durations, as long as the sidebands are still resolved. 

\begin{table}[]
    \centering
    \begin{tabular}{c |c| c}
        $n$ &T($\lvert n \rangle$) ($\mu$s) & T($\lvert 0 \rangle + \lvert n \rangle$) ($\mu$s) \\
        2 & 22 & 16 \\
        4 & 38 & 41 \\
        6 & 51 & 54 \\
        8 & 63 & 66 \\
        12 & 86 & 89 \\
        16 & 109 & 112 \\
        20 & 133 & -- \\
        40 & 196 & -- \\
        80 & 293 & -- \\
        100 & 335 & -- \\
    \end{tabular}
    \caption{Duration of pulse sequences
    to produce number states and number-state superpositions, with a base Rabi frequency $\Omega_{0,1}=\pi/(13~\mu$s). Beyond $n=40$, we make use of higher-order sidebands, which allows us to use transitions with higher Rabi frequencies and skip ``rungs" as we move the ion up the number-state ladder. }
    \label{Tab:exptimes}
\end{table}
The sidebands in our interferometers rely on the same Raman laser coupling that is used in most two-qubit gates, so the time scale for a single BSB or RSB pulse is on the same order as a typical gate time. This limits the usefulness for improving gate fidelity, but interferometric tracking would certainly help with longer-term drifts and reduce the duration required to measure the trap frequency with a certain precision. 
\subsubsection*{Sources of decoherence}
At the 7.2 MHz axial mode frequency, the heating rate is relatively low, approximately 17 quanta/s. Consequently, the dominant source of decoherence in these experiments is dephasing due to fluctuations in the mode frequency. These fluctuations probably arise from technical noise from the voltage source, 
uncontrolled charging from stray light scattering off the dielectric material between the trap electrodes and amplitude instabilities in the rf source that generates the pseudo-potential. We also attribute longer time-scale drift, on the order of minutes to hours and a magnitude of $\sim 1 - 10$ Hz/s to uncontrolled charging and discharging. This charging affects mode-frequency tracking experiments and leads to unpredictable deviations from the ideal white-noise $1/\sqrt{\tau}$ behavior during experimental runs to determine Allan-variances. A more detailed investigation of this noise and ways to further suppress it is currently in progress in our laboratory. 
\subsubsection*{Auto-balanced frequency tracking experiments}
\begin{figure}
\centering
\includegraphics{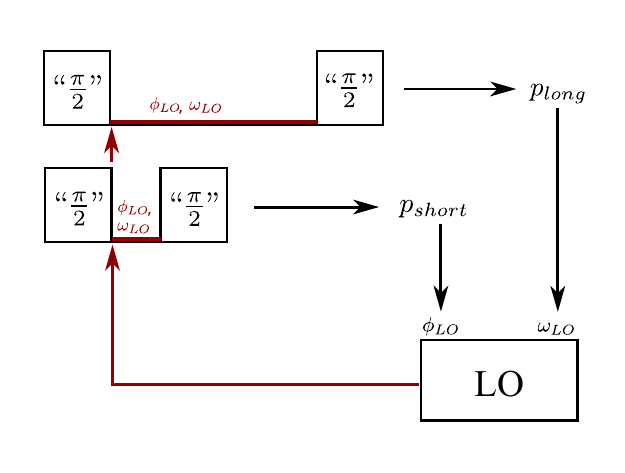}
\caption{{\bf Schematic illustrating the auto-balanced feedback loop applied to local oscillator (LO), a frequency source used as a reference to compare to the ion's oscillation frequency.} The LO controls the phases and frequencies of the BSB and RSB laser-pulses (see Fig. \ref{Fig:numberstate} a)) during mode-frequency tracking experiments. The difference between the populations measured after a pair of Ramsey experiments with long wait times provides an error signal, $p_{long}$, used to feed back on the LO frequency, $\omega_{LO}$. Similarly, a 
second pair of Ramsey experiments with short wait times provides and error signal, $p_{short}$, used to feed back on an additional LO phase offset $\phi_{LO}$ between the first and second effective $\pi/2$-pulses. The long and short Ramsey experiments are interleaved, with $\phi_{LO}$ and $\omega_{LO}$ applied equally to both. For more details on auto-balanced Ramsey experiments, see Ref. \citeonline{Sann17}.
    }
    \label{Fig:trackingschem}
\end{figure}

The auto-balanced sequence (see Fig. \ref{Fig:trackingschem}) for mode-frequency tracking is comprised of four interleaved Ramsey experiments, two each with Ramsey times $t_{short}$ and $t_{long}$, where the phase of the second effective $\pi/2$-pulse is $+\pi/2$ relative to the first $\pi/2$-pulse for one of the experiments and $-\pi/2$ for the other. For a given Ramsey time, the two  experiments with $+\pi/2$ and $-\pi/2$ relative phase interrogate the fringe close to its largest positive and negative slope respectively. If the long Ramsey experiments are on exact resonance, both should result in the same average population in $\lvert \downarrow\rangle$, so the signal difference is zero on average. If the Ramsey experiment is off resonance, the signal difference provides a non-zero error signal $p_{long}$ which is used to calculate the offset $\delta\omega$ between the assumed mode frequency $\omega_a$ and the actual mode frequency. The offset $\delta \omega$ is fed back to the local oscillator (LO), which updates its frequency to $\omega=\omega_a+\delta \omega$. This in turn updates the frequency of the BSB and RSB pulses to $\omega_{LO} = \omega_0 + \omega_a + \delta\omega$ and $\omega_{LO} = \omega_0 - \omega_a - \delta\omega$, respectively, where $\hbar \omega_0$ is the energy difference between $\lvert\uparrow\rangle$ and $\lvert\downarrow\rangle$, as described in the main text. The short Ramsey experiments provide an error signal $p_{short}$, as described above, which is now used to compute a phase offset $\delta\phi$ that is added to the relative phase between the two effective $\pi$-pulses, $\phi_{LO} = \pm\pi/2+\delta\phi$. Feedback on this phase reduces ``frequency pulling" due to non-zero phase accumulation during the pulses, which can be caused by slowly drifting, pulse-synchronous systematic errors such as AC-Stark shifts \cite{Sann17}. For a more detailed description of the auto-balanced Ramsey technique, see Ref. \citeonline{Sann17}.

\subsubsection*{Anharmonic contributions}
We do not expect the anharmonicity of the trapping potential to be a significant limitation or source of systematic error. Previous calculations\cite{Home11} for a similar trap and estimates based on a numerical simulation of our current trap predict an anharmonic component of a few parts in $10^{-7}$ per quantum on the axial mode frequency. With a maximal superposition state of $\lvert 0 \rangle + \lvert 8 \rangle$ used in the mode-tracking experiments, this would cause an offset of $\sim 1 \times 10^{-6}$, which is within the minimum measurement uncertainty. \\
\subsection*{Data availability}
The data sets generated during and/or analyzed during the current study are available from the corresponding author on reasonable request.
\end{document}